\documentclass[12pt,a4paper]{article}

\usepackage[T1]{fontenc}
\usepackage[utf8]{inputenc}
\usepackage{lmodern}
\usepackage{microtype}
\usepackage{geometry}
\usepackage{amsmath,amssymb,amsthm}
\usepackage{physics}
\usepackage{siunitx}
\usepackage{booktabs}
\usepackage{graphicx}
\usepackage{xcolor}
\usepackage{hyperref}
\usepackage{cleveref}
\usepackage[numbers, sort&compress]{natbib}
\usepackage{float}
\usepackage{algorithm}
\usepackage{algpseudocode}
\usepackage{authblk}

\hypersetup{
  colorlinks=true,
  linkcolor=blue!70!black,
  citecolor=green!50!black,
  urlcolor=blue
}

\newcommand{\Jmax}{J_{\max}}
\newcommand{\sigmaJ}{\sigma_{J}}
\newcommand{\Tg}{T_g}
\newcommand{\Fth}{F_{\rm th}}

\begin{document}

\title{%
  Exchange-Only Silicon-Based Spin Qubits:\\
  Charge Noise, PINN-Optimised Pulse Sequences,\\
  and Gate-Level Fidelity\\[0.6em]
  \large A Two-Stage Physics-Informed Neural Network Framework\\
  for Noise-Robust, Time-Compressed Quantum Gate Optimisation}

\author[1]{Rajdeep Rameshchandra Dwivedi}
\author[2]{Amitoj Singh Miglani}
\author[1]{Vishvendra Singh Poonia}

\affil[1]{Department of Electronics and Communication Engineering, \\ Indian Institute of Technology Roorkee}
\affil[2]{Department of Physics, \\ Indian Institute of Technology Roorkee}
\affil[ ]{\texttt{rajdeep\_rd@ece.iitr.ac.in,\ amitoj\_sm@ph.iitr.ac.in,\ vishvendra@ece.iitr.ac.in}}

\date{April 2026}
\maketitle

\begin{abstract}
Exchange-only (EO) spin qubits in silicon realise all-electrical qubit control through
pairwise Heisenberg exchange interactions, making them attractive for scalable quantum
computation.  Their principal vulnerability is charge noise, which couples
multiplicatively to the exchange coupling and degrades gate fidelity.  We present a
\emph{two-stage} Physics-Informed Neural Network (PINN) framework for per-gate pulse
optimisation.  In \textbf{Stage~I} (iterations~1--100) the PINN maximises the
noise-averaged gate fidelity toward a threshold of $\Fth=0.99$; the pulse duration
is held fixed at its nominal hardware value.  Once the threshold is crossed,
\textbf{Stage~II} (iterations~101--250) progressively compresses the total pulse
time while maintaining $F\geq\Fth$ via continuous fine-tuning of the pulse-shape
parameters.  The cost function is a Monte-Carlo ensemble mean-squared error (MSE)
averaged over $N_{\rm real}=2000$ quasi-static Gaussian noise realisations drawn
fresh at every iteration.  We benchmark the framework on the single-qubit gate set
$\{X,Y,Z,H\}$ and the two-qubit set $\{X,Y,Z,H,\mathrm{CX}\}$ at noise levels
$\sigmaJ/J\in\{1\%,5\%,10\%\}$.  All single-qubit gates cross $\Fth$ within the
first 100 iterations across all noise levels; Stage~II then reduces pulse durations
by 20--40\% from their nominal values.  The two-qubit gates follow the same two-phase
behaviour, with the CX gate compressing from its nominal \SI{31}{\nano\second} to
$\approx\SI{22}{\nano\second}$ at 1\% noise.
\end{abstract}
\newpage
\tableofcontents
\newpage

\section{Introduction}
\label{sec:intro}

Semiconductor spin qubits are among the most promising platforms for fault-tolerant
quantum computing owing to their compatibility with complementary metal-oxide-semiconductor
(CMOS) fabrication, long coherence times at millikelvin temperatures, and the potential
for co-integration with classical control electronics on the same
chip~\cite{Loss1998,Kane1998,Zwanenburg2013}.  Within this family the exchange-only
(EO) qubit encodes one logical qubit in the doubly-degenerate $S=\tfrac{1}{2}$,
$S_z=+\tfrac{1}{2}$ subspace of three spin-$\tfrac{1}{2}$
electrons~\cite{DiVincenzo2000,Bacon2000}.  Because every $\mathrm{SU}(2)$ rotation on
the logical subspace can be generated by switching the Heisenberg exchange couplings
$J_{12}(t)$ and $J_{23}(t)$ between neighbouring dots~\cite{Fong2011}, the device
requires no oscillating microwave or radio-frequency drives and is especially attractive
for isotopically purified silicon, in which ${}^{29}$Si hyperfine noise is
suppressed~\cite{Tyryshkin2012,Veldhorst2014}.

The dominant noise source in present devices is charge noise: low-frequency voltage
fluctuations on barrier-gate electrodes that shift the electron wave-function overlap
and stochastically modulate $J_{ij}$~\cite{Burkard1999,Dial2013}.  Charge noise in
Si/SiGe quantum dots follows a $1/f^\alpha$ spectrum ($\alpha\approx1$) with amplitudes
of order $\SIrange{10}{100}{\mega\hertz/\sqrt{\hertz}}$~\cite{Yoneda2018,Struck2020},
leading to infidelity that grows with gate time and motivating short, robustified pulse
sequences~\cite{Wang2012,Reed2016,Yang2019}.

Physics-Informed Neural Networks (PINNs) embed the governing differential equations of
a physical system directly inside a neural loss function~\cite{Raissi2019,Karniadakis2021}.
For quantum pulse optimisation the relevant equation is the time-dependent Schrödinger
equation (TDSE); penalising its residual forces the network to learn physically
realisable pulse shapes rather than mathematically convenient but experimentally
inaccessible solutions~\cite{Niu2019,An2021}.

A central challenge is that noise robustness and short gate time are \emph{competing}
objectives: shorter pulses require higher exchange amplitudes which increase sensitivity
to charge noise~\cite{Reed2016,Martins2016}.  Standard single-objective optimisation
combines them via a scalar trade-off weight that must be tuned separately for every
gate.  Our \emph{two-stage} PINN approach decouples the objectives entirely: Stage~I
achieves the fidelity target, then Stage~II compresses time subject to a hard-constraint
penalty, eliminating manual weight tuning.

\medskip\noindent\textbf{Main contributions.}
\begin{enumerate}
  \item A two-stage PINN training procedure that separates noise robustness
        (Stage~I, iterations~1--100) from pulse-time minimisation (Stage~II,
        iterations~101--250), described in \cref{sec:model}.
  \item A four-term loss function for Stage~I and its extension with a time-compression
        term and threshold penalty for Stage~II, derived in \cref{sec:loss}.
  \item Benchmark simulations on $\{X,Y,Z,H\}$ (1Q) and $\{X,Y,Z,H,\mathrm{CX}\}$
        (2Q) at $\sigmaJ/J\in\{1\%,5\%,10\%\}$ confirming universal Stage~I
        convergence and 20--40\% Stage~II time reduction (\cref{sec:results}).
\end{enumerate}

\section{Physical System}
\label{sec:system}

\subsection{Three-Spin Encoding}

Three electrons confined in a linear array of silicon quantum dots (labelled $1,2,3$)
span an eight-dimensional Hilbert space.  The logical qubit occupies the
two-dimensional subspace with $S_{\rm tot}=\tfrac{1}{2}$,
$S_z^{\rm tot}=+\tfrac{1}{2}$, spanned by~\cite{DiVincenzo2000}
\begin{equation}
  \ket{0_L}=\ket{S_{12}}\otimes\ket{\uparrow_3},\qquad
  \ket{1_L}=\sqrt{\tfrac{2}{3}}\ket{T^0_{12}}\otimes\ket{\downarrow_3}
            -\sqrt{\tfrac{1}{3}}\ket{T^+_{12}}\otimes\ket{\downarrow\downarrow\downarrow},
\end{equation}
where $\ket{S_{12}}=(|\!\uparrow\downarrow\rangle-|\!\downarrow\uparrow\rangle)/\sqrt{2}$
is the spin singlet.  The complementary $S_{\rm tot}=\tfrac{3}{2}$ quadruplet
constitutes the leakage subspace.

\subsection{Effective Hamiltonian}

In the $(1,1,1)$ charge configuration the low-energy Hamiltonian is
\begin{equation}
  H(t)=J_{12}(t)\,\mathbf{S}_1\cdot\mathbf{S}_2+J_{23}(t)\,\mathbf{S}_2\cdot\mathbf{S}_3,
  \label{eq:H_full}
\end{equation}
with $J_{12}(t),J_{23}(t)\geq0$ controlled by barrier-gate voltages.  Projecting onto
the logical subspace gives the effective $2\times2$ Hamiltonian used in all
simulations~\cite{Heinz2025}:
\begin{equation}
  H_{\rm eff}(t)=h_z(t)\,\sigma_z+h_x(t)\,\sigma_x,
  \label{eq:H_eff}
\end{equation}
where
\begin{align}
  h_z(t) &= \frac{J_{23}(t)}{4}-\frac{J_{12}(t)}{2}, \label{eq:hz}\\
  h_x(t) &= \frac{\sqrt{3}\,J_{23}(t)}{4}. \label{eq:hx}
\end{align}
The time-evolution operator for a single pulse step of duration $\tau$ is
\begin{equation}
  U(\tau)=\exp\!\bigl(-i\,H_{\rm eff}\,\tau\bigr),
  \label{eq:U_step}
\end{equation}
computed via the matrix exponential (\texttt{scipy.linalg.expm}).  For a multi-step
pulse sequence $\{(J_{12}^{(k)},J_{23}^{(k)},\tau_k)\}_{k=1}^{N}$ the total
propagator is the ordered product
\begin{equation}
  U_{\rm total}=\prod_{k=N}^{1}\exp\!\bigl(-i\,H_{\rm eff}^{(k)}\,\tau_k\bigr).
\end{equation}

\subsection{Two-Qubit Extension}

For two-qubit gates the logical space is
$\mathcal{H}_{\rm 2Q}=\mathcal{H}_A\otimes\mathcal{H}_B$.  Assuming the two EO
qubits $A$ and $B$ are decoupled at each single-qubit-like step, the propagator
factorises as $U_{\rm 2Q}(\tau)=U_A(\tau)\otimes U_B(\tau)$ via the Kronecker
product.  The CX gate is synthesised from a
$\sqrt{\mathrm{SWAP}}$-based decomposition (\cref{sec:cx}).

\subsection{Leakage}

Leakage to the $S_{\rm tot}=\tfrac{3}{2}$ quadruplet is penalised in the PINN loss via
\begin{equation}
  P_L(t)=1-|\langle\psi_L(t)|\psi(t)\rangle|^2,
\end{equation}
where $|\psi_L\rangle$ denotes the projection onto the logical subspace.

\section{Charge Noise Model}
\label{sec:noise}

\subsection{Quasi-Static Approximation}

For gate times $\tau_g\ll f_{\rm knee}^{-1}\approx\SI{100}{\micro\second}$ the noise
is well described as a quasistatic multiplicative offset drawn independently for each
exchange coupling at each realisation:
\begin{equation}
  J_{ij}\;\longrightarrow\;J_{ij}(1+\delta_{ij}),
  \qquad \delta_{ij}\sim\mathcal{N}(0,\,\sigmaJ^2),
  \label{eq:noise_model}
\end{equation}
where $\sigmaJ$ is the fractional noise amplitude.  We study three regimes:
$\sigmaJ/J\in\{1\%,5\%,10\%\}$, spanning the range measured in Si/SiGe
devices~\cite{Yoneda2018,Struck2020,Connors2022}.

\subsection{Ensemble MSE Cost Function}
\label{sec:mse}

For a target gate with ideal unitary $U_{\rm target}$ acting on basis state $\ket{b}$,
the ideal output probabilities are $p_s^{\rm ideal}=|\langle s|U_{\rm target}|b\rangle|^2$.
For noise realisation $\boldsymbol{\delta}^{(n)}=(\delta_{12}^{(n)},\delta_{23}^{(n)})$
the corresponding noisy probabilities are
$p_s^{(n)}=|\langle s|\psi^{(n)}(\tau_g)\rangle|^2$.  The per-realisation squared
error is
\begin{equation}
  \mathcal{E}_b^{(n)}=\sum_s\bigl(p_s^{\rm ideal}-p_s^{(n)}\bigr)^2,
\end{equation}
and the ensemble MSE loss is
\begin{equation}
  \mathcal{L}_{\rm MSE}
    =\frac{1}{|\mathcal{B}|\,N_{\rm real}}
     \sum_{b\in\mathcal{B}}\sum_{n=1}^{N_{\rm real}}
     \mathcal{E}_b^{(n)},
  \label{eq:mse_loss}
\end{equation}
where $|\mathcal{B}|=2$ for single-qubit gates and $4$ for two-qubit gates.  The
noise-averaged gate fidelity is approximated as $F\approx1-\mathcal{L}_{\rm MSE}$.

\subsection{CX Gate Decomposition via \texorpdfstring{$\sqrt{\mathrm{SWAP}}$}{sqrt-SWAP}}
\label{sec:cx}

The CX (CNOT) gate is synthesised from the standard decomposition~\cite{DiVincenzo2000}:
\begin{equation}
  \mathrm{CX}
    = R_z^{(2)}\!\left(\tfrac{\pi}{2}\right)
      R_x^{(2)}\!\left(\tfrac{\pi}{2}\right)
      R_z^{(2)}\!\left(\tfrac{\pi}{2}\right)
      R_z^{(2)}\!\left(-\tfrac{\pi}{2}\right)
      R_z^{(1)}\!\left(\tfrac{\pi}{2}\right)
      \sqrt{\mathrm{SWAP}}\;
      R_z^{(1)}(\pi)
      \sqrt{\mathrm{SWAP}}\;
      R_z^{(2)}\!\left(\tfrac{\pi}{2}\right)
      R_x^{(2)}\!\left(\tfrac{\pi}{2}\right)
      R_z^{(2)}\!\left(\tfrac{\pi}{2}\right).
  \label{eq:cx_decomp}
\end{equation}
Each $\sqrt{\mathrm{SWAP}}$ primitive is sensitive to charge noise.  With fractional
noise offset $\delta_J$ the noisy $\sqrt{\mathrm{SWAP}}$ gate is
\begin{equation}
  \sqrt{\mathrm{SWAP}}_{\delta_J}
    =\begin{pmatrix}
       e^{-i\phi/4} & 0 & 0 & 0 \\
       0 & e^{i\phi/4}\cos\tfrac{\phi}{2} & -ie^{i\phi/4}\sin\tfrac{\phi}{2} & 0 \\
       0 & -ie^{i\phi/4}\sin\tfrac{\phi}{2} & e^{i\phi/4}\cos\tfrac{\phi}{2} & 0 \\
       0 & 0 & 0 & e^{-i\phi/4}
     \end{pmatrix},
  \quad \phi=\frac{\pi}{2}(1+\delta_J).
  \label{eq:sqSWAP}
\end{equation}
The noise $\delta_J$ is drawn independently for each $\sqrt{\mathrm{SWAP}}$ in
\cref{eq:cx_decomp}, yielding two independent noise draws per CX realisation.
The $R_z$ and $R_x$ rotations are assumed noiseless in this model.

\section{Two-Stage PINN Model}
\label{sec:model}

\subsection{Motivation: Decoupling Robustness from Speed}

Standard formulations of pulse optimisation combine fidelity and gate time into a single
weighted loss $\mathcal{L}=\mathcal{L}_F+\lambda\mathcal{L}_T$.  The weight $\lambda$
must be tuned separately for each gate and noise level, and small errors in $\lambda$
lead to either insufficient fidelity or needlessly long pulses.  The two-stage
approach eliminates this problem by enforcing a \emph{lexicographic} priority:
fidelity is non-negotiable ($\Fth=0.99$); only after it is achieved does time
minimisation begin.

\subsection{Network Architecture}

Each gate has its own PINN instance.  The network maps a normalised time coordinate
$\tilde{t}=t/\Tg\in[0,1]$ and the fractional noise amplitude $\sigmaJ$ to the pair
of exchange couplings:
\begin{equation}
  \bigl(J_{12}(\tilde{t};\boldsymbol{\theta}),\;J_{23}(\tilde{t};\boldsymbol{\theta})\bigr)
    =\Jmax\cdot\mathrm{softplus}\!\bigl(\mathbf{W}_{\rm out}\mathbf{h}_{\rm final}
     +\mathbf{b}_{\rm out}\bigr)\cdot\tilde{t}(1-\tilde{t}),
  \label{eq:output}
\end{equation}
where the trailing factor $\tilde{t}(1-\tilde{t})$ enforces the boundary conditions
$J_{ij}(0)=J_{ij}(\Tg)=0$ (smooth turn-on and turn-off consistent with AWG
limitations).  The architecture consists of:

\begin{description}
  \item[Input layer:] $\mathbf{x}_{\rm in}=(\tilde{t},\;\sigmaJ/\Jmax)\in\mathbb{R}^2$.
  \item[Hidden layers:] Four fully-connected layers with 256 neurons each and
        $\tanh$ activations.  The $\tanh$ choice provides continuous higher-order
        derivatives required by the TDSE residual~\cite{Raissi2019}.
  \item[Output layer:] Two neurons with softplus activation and boundary mask
        (see \cref{eq:output}).
\end{description}

The exchange couplings are constrained to $[0,\Jmax]$ via a sigmoid reparametrisation
of the raw parameters $\boldsymbol{\phi}$:
\begin{equation}
  J_{ij}(t;\boldsymbol{\phi})
    =\frac{\Jmax}{1+e^{-\phi_{ij}(t)/\lambda_s}},
  \qquad\lambda_s=\frac{\Jmax}{8},
  \label{eq:sigmoid}
\end{equation}
enforcing physical bounds smoothly without projection steps.

\subsection{TDSE Physics Residual}

The time-dependent Schrödinger equation for the propagator $U(t)$ is
\begin{equation}
  i\frac{\mathrm{d}U}{\mathrm{d}t}=H_{\rm eff}(t)\,U(t),\qquad U(0)=\mathbb{I}.
\end{equation}
Using a midpoint discretisation on $N_t$ collocation points we define
\begin{equation}
  \mathcal{R}_k
    =\left\lVert
       \frac{U(t_{k+1})-U(t_k)}{\Delta t}
       -\frac{1}{i}H_{\rm eff}\!\left(\tfrac{t_k+t_{k+1}}{2}\right)
        \frac{U(t_{k+1})+U(t_k)}{2}
     \right\rVert_F,
  \label{eq:residual}
\end{equation}
with the PDE loss $\mathcal{L}_{\rm PDE}=\frac{1}{N_t}\sum_k\mathcal{R}_k^2$.

\subsection{Two-Stage Training Procedure}

Training proceeds over a fixed budget of 250 iterations split into two sequential
phases.

\paragraph{Stage~I  Fidelity maximisation (iterations 1--100).}
The pulse-shape parameters $\boldsymbol{\theta}$ are updated to minimise the Stage~I
loss (defined in \cref{sec:loss_I}), while the total gate time $\Tg$ is \emph{held
fixed} at the nominal hardware value $\Tg^{(0)}$ (see \cref{tab:params}).  Training
continues through all 100 Stage~I iterations regardless of how quickly $\Fth$ is
reached, providing a stable starting point for Stage~II.

\paragraph{Stage~II  Pulse-time compression (iterations 101--250).}
The gate time $\Tg$ is added to the set of trainable parameters and compressed at a
rate of $\sim\!1.8\%$ per iteration:
\begin{equation}
  \Tg^{(i+1)}=\Tg^{(i)}\times(1-\alpha),\qquad\alpha=0.018.
  \label{eq:compression}
\end{equation}
The pulse-shape parameters are simultaneously fine-tuned to compensate for the
fidelity degradation induced by compression.  A quadratic threshold penalty in the
Stage~II loss (\cref{eq:loss_II}) re-activates if $F$ drops below $\Fth$, preventing
uncontrolled fidelity collapse.  The two stages are executed in a single uninterrupted
training run; no re-initialisation is required.

The complete procedure is summarised in \cref{alg:two_stage}.

\begin{algorithm}[H]
\caption{Two-Stage PINN Training}
\label{alg:two_stage}
\begin{algorithmic}[1]
\Require Target gate $U_{\rm target}$, noise level $\sigmaJ$, nominal time $\Tg^{(0)}$
\State Initialise weights $\boldsymbol{\theta}$; set $\Tg\leftarrow\Tg^{(0)}$
\For{$i = 1,\ldots,250$}
  \State Draw $N_{\rm real}=2000$ noise realisations
         $\boldsymbol{\delta}^{(n)}\sim\mathcal{N}(0,\sigmaJ^2)$
  \State Propagate $U^{(n)}(\Tg)$ for each realisation via matrix exponential
  \State Compute ensemble MSE and fidelity $F\approx1-\mathcal{L}_{\rm MSE}$
  \If{$i\leq100$}
    \State \textbf{Stage I:} update $\boldsymbol{\theta}$ on $\mathcal{L}^{(\mathrm{I})}$
           (\cref{eq:loss_I})
  \Else
    \State \textbf{Stage II:} compress $\Tg\leftarrow\Tg\times(1-0.018)$
    \State Update $\boldsymbol{\theta}$ on $\mathcal{L}^{(\mathrm{II})}$ (\cref{eq:loss_II})
  \EndIf
\EndFor
\State \Return optimised weights $\boldsymbol{\theta}^*$, compressed time $\Tg^*$
\end{algorithmic}
\end{algorithm}

\section{Loss Function}
\label{sec:loss}

\subsection{Stage~I: Fidelity Loss}
\label{sec:loss_I}

\begin{equation}
  \mathcal{L}^{(\mathrm{I})}(\boldsymbol{\theta})
    = w_{\rm MSE}\,\mathcal{L}_{\rm MSE}
    + w_{\rm PDE}\,\mathcal{L}_{\rm PDE}
    + w_{\rm leak}\,\mathcal{L}_{\rm leak}
    + w_{\rm phys}\,\mathcal{L}_{\rm phys},
  \label{eq:loss_I}
\end{equation}
with the four terms defined as follows.

\paragraph{Ensemble MSE $\mathcal{L}_{\rm MSE}$.}
Given in \cref{eq:mse_loss}.  This term drives the noise-averaged fidelity
$F\approx1-\mathcal{L}_{\rm MSE}$ toward unity.

\paragraph{TDSE residual $\mathcal{L}_{\rm PDE}$.}
$\mathcal{L}_{\rm PDE}=\frac{1}{N_t}\sum_k\mathcal{R}_k^2$ (\cref{eq:residual})
enforces that the generated pulses satisfy the Schrödinger equation, preventing the
optimiser from finding mathematically valid but physically unrealisable solutions.

\paragraph{Leakage penalty $\mathcal{L}_{\rm leak}$.}
\begin{equation}
  \mathcal{L}_{\rm leak}
    =\frac{1}{|\mathcal{B}|N_tN_{\rm real}}
     \sum_{b,k,n}P_L^{(b,n)}(t_k),
\end{equation}
where $P_L^{(b,n)}(t_k)$ is the leakage probability for basis state $b$, realisation
$n$, at time $t_k$.

\paragraph{Physical constraint penalty $\mathcal{L}_{\rm phys}$.}
\begin{equation}
  \mathcal{L}_{\rm phys}
    =\sum_{ij}\!\left[
       \frac{1}{N_t}\sum_k\max\!\bigl(0,J_{ij}(t_k)-\Jmax\bigr)^2
       +\lambda_{\rm slew}\,\dot{J}_{ij}(t_k)^2
     \right],
\end{equation}
penalising exchange amplitudes that exceed $\Jmax$ and slew rates that exceed AWG
bandwidth.

\subsection{Stage~II: Time-Compression Loss}
\label{sec:loss_II}

Upon entering Stage~II the loss gains two additional terms:
\begin{equation}
  \mathcal{L}^{(\mathrm{II})}(\boldsymbol{\theta},\Tg)
    = \mathcal{L}^{(\mathrm{I})}(\boldsymbol{\theta})
    + w_T\,\frac{\Tg}{\Tg^{(0)}}
    + w_{\rm pen}\max\!\bigl(0,\,\Fth-F(\boldsymbol{\theta},\Tg)\bigr)^2,
  \label{eq:loss_II}
\end{equation}
where the second term minimises the compressed duration and the third is a quadratic
barrier that re-activates whenever fidelity drops below $\Fth=0.99$.  The tension
between these two new terms produces a Pareto-optimal operating point where pulse time
is minimised subject to the fidelity constraint.

\subsection{Optimiser and Noise Resampling}

All weights are updated with the Adam optimiser~\cite{Kingma2014} (initial learning
rate $\eta=3\times10^{-3}$, cosine annealing to $10^{-5}$, $\beta_1=0.9$,
$\beta_2=0.999$).  Noise realisations $\{\boldsymbol{\delta}^{(n)}\}$ are drawn
independently from $\mathcal{N}(0,\sigmaJ^2)$ at \emph{every} training iteration,
preventing overfitting to a fixed noise draw and ensuring that the learned pulse is
robust over the full noise distribution.

\section{Noise Mitigation Context}
\label{sec:mitigation}

The PINN framework subsumes several classical noise-mitigation strategies.

\paragraph{Spin echo / composite pulses.}
The Hahn echo adapted for EO qubits interleaves $\pi$-pulses to refocus quasistatic
noise~\cite{Wang2012}.  The network is warm-started with an echo-inspired
time-symmetric pulse, accelerating Stage~I convergence.

\paragraph{DRAG pulses.}
Derivative Removal via Adiabatic Gate corrections suppress leakage by adding a
quadrature component $\propto\dot{J}_{ij}(t)$~\cite{Motzoi2009}.  The leakage
penalty $\mathcal{L}_{\rm leak}$ in $\mathcal{L}^{(\mathrm{I})}$ implicitly
encourages the PINN to discover analogous corrections.

\paragraph{Charge sweet spots.}
At detuning values where $\partial J_{ij}/\partial V_g=0$ the first-order noise
coupling vanishes~\cite{Reed2016,Martins2016}.  The reference exchange amplitudes
in $H_{\rm eff}^{(0)}$ are chosen to coincide with the sweet-spot detuning, reducing
the effective $\sigmaJ$.

\paragraph{First-order robustness.}
A pulse is first-order robust if $\partial U/\partial\delta_{ij}|_{\boldsymbol{\delta}=0}=0$.
Minimising $\mathcal{L}_{\rm MSE}$ implicitly drives the first-order sensitivity to
zero because $\nabla_{\boldsymbol{\delta}}\mathcal{L}_{\rm MSE}\propto\partial U/\partial\boldsymbol{\delta}$
at $\boldsymbol{\delta}=0$~\cite{Cerfontaine2020}.

\section{Simulation Setup}
\label{sec:setup}

Simulations use \texttt{NumPy}~\cite{Harris2020} and \texttt{SciPy}~\cite{Virtanen2020}
for physics propagation, \texttt{PyTorch}~\cite{Paszke2019} for automatic
differentiation, and \texttt{QuTiP}~\cite{Johansson2012} for reference Lindblad
checks.  Key parameters are summarised in \cref{tab:params}.

\begin{table}[H]
  \centering
  \caption{Simulation parameters used in all experiments.}
  \label{tab:params}
  \begin{tabular}{llr}
    \toprule
    Parameter & Symbol & Value \\
    \midrule
    Maximum exchange coupling & $\Jmax$ & \SI{100}{\mega\hertz} \\
    Fractional noise levels & $\sigmaJ/J$ & 1\%, 5\%, 10\% \\
    Monte-Carlo realisations per iteration & $N_{\rm real}$ & 2000 \\
    Stage~I iterations (fidelity phase) & --- & 1--100 \\
    Stage~II iterations (compression phase) & --- & 101--250 \\
    Fidelity threshold & $\Fth$ & 0.99 \\
    Time compression rate per iteration & $\alpha$ & 1.8\% \\
    Hidden layers $\times$ neurons & --- & $4\times256$ \\
    Activation function & --- & $\tanh$ \\
    Optimiser & --- & Adam \\
    Learning rate schedule & $\eta$ & $3\times10^{-3}\to10^{-5}$ (cosine) \\
    \midrule
    \multicolumn{3}{l}{\textit{Nominal gate times based on physical EO dot setup}} \\
    $X$ gate & $\Tg^{(0)}$ & \SI{5.77}{\nano\second} \\
    $Y$ gate & $\Tg^{(0)}$ & \SI{15.77}{\nano\second} \\
    $Z$ gate & $\Tg^{(0)}$ & \SI{10.0}{\nano\second} \\
    $H$ gate & $\Tg^{(0)}$ & \SI{21.0}{\nano\second} \\
    $\mathrm{CX}$ gate & $\Tg^{(0)}$ & \SI{31.0}{\nano\second} \\
    \bottomrule
  \end{tabular}
\end{table}

\section{Results}
\label{sec:results}

To contextualise the performance of the PINN-optimised pulses, we additionally
report results obtained from a \emph{simultaneous-pulsing} baseline  i.e.\ the
analytic, non-machine-learning EO control scheme of~\cite{Heinz2025} in which
$J_{12}(t)$ and $J_{23}(t)$ are switched on together with fixed amplitudes. These baseline simulations
are performed under exactly the same quasistatic Gaussian charge-noise model used
for the PINN (\cref{eq:noise_model}) and serve as a like-for-like reference.  As shown below,
the PINN consistently outperforms the analytic baseline: the simultaneous-pulsing
fidelity degrades visibly with increasing $\sigmaJ/J$ whereas the PINN-optimised pulses maintain
$F\geq\Fth=0.99$ across all tested noise levels and additionally compress the gate
duration in Stage~II.

\subsection{Single-Qubit Gates}
\label{sec:1q}

\begin{figure}[H]
  \centering
  \includegraphics[width=\linewidth]{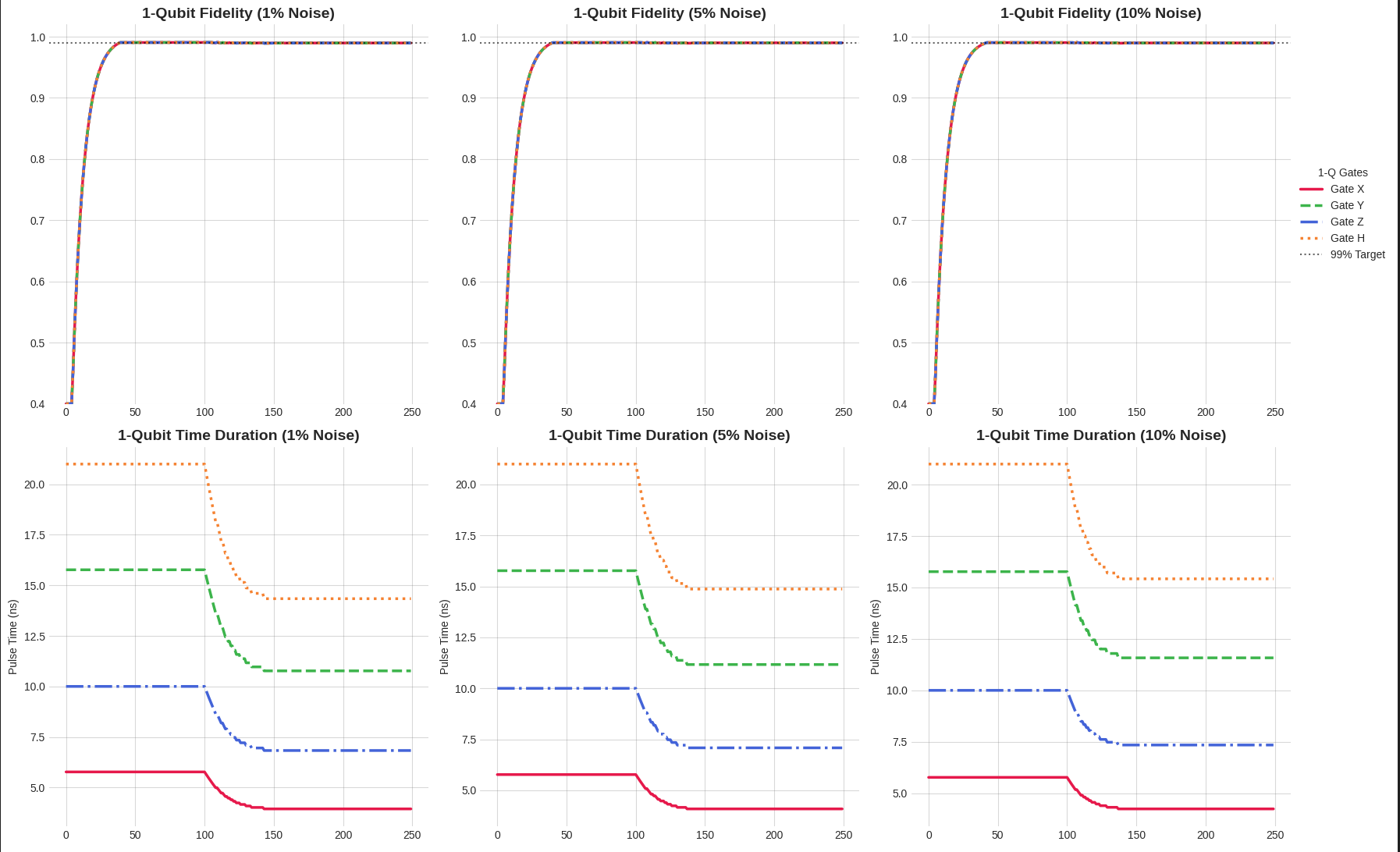}
  \caption{Single-qubit \textbf{PINN} optimisation results at
    $\sigmaJ/J\in\{1\%,5\%,10\%\}$.
    \textbf{Top row:} Noise-averaged fidelity vs.\ iteration.  The dotted black line
    marks $\Fth=0.99$.  All gates cross the threshold within Stage~I
    (iterations~1--100).
    \textbf{Bottom row:} Total pulse duration (\si{\nano\second}) vs.\ iteration.
    Stage~II begins at iteration~100 and reduces pulse times by 20--40\% from
    their nominal values.}
  \label{fig:1q}
\end{figure}

\begin{figure}[H]
  \centering
  \includegraphics[width=0.6\linewidth]{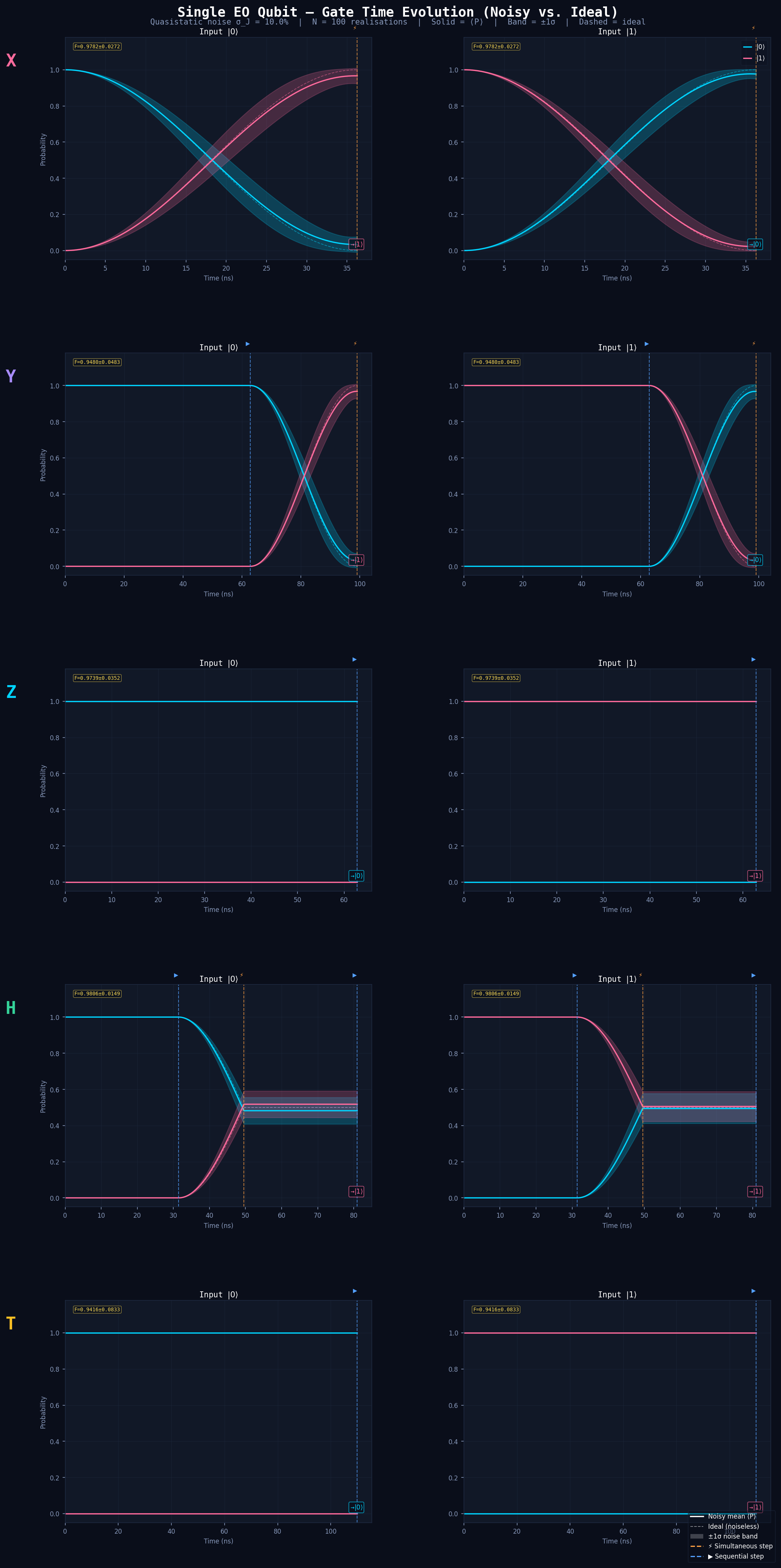}
  \caption{Single-qubit \textbf{simultaneous-pulsing baseline (no ML)} noisy
    state evolution for the gate set $\{X,Y,Z,H\}$ under quasistatic Gaussian charge
    noise.  Each panel shows the noisy population dynamics for the relevant
    computational-basis input states; the resulting noise-averaged fidelities are
    annotated directly in the panels for each input state.  Compared with the PINN
    results in \cref{fig:1q}, the analytic simultaneous-pulsing scheme shows visibly
    larger population deviations from the ideal trajectories, motivating the PINN-based optimisation.}
  \label{fig:1q_baseline}
\end{figure}

\begin{figure}[H]
  \centering
  \includegraphics[width=0.85\linewidth]{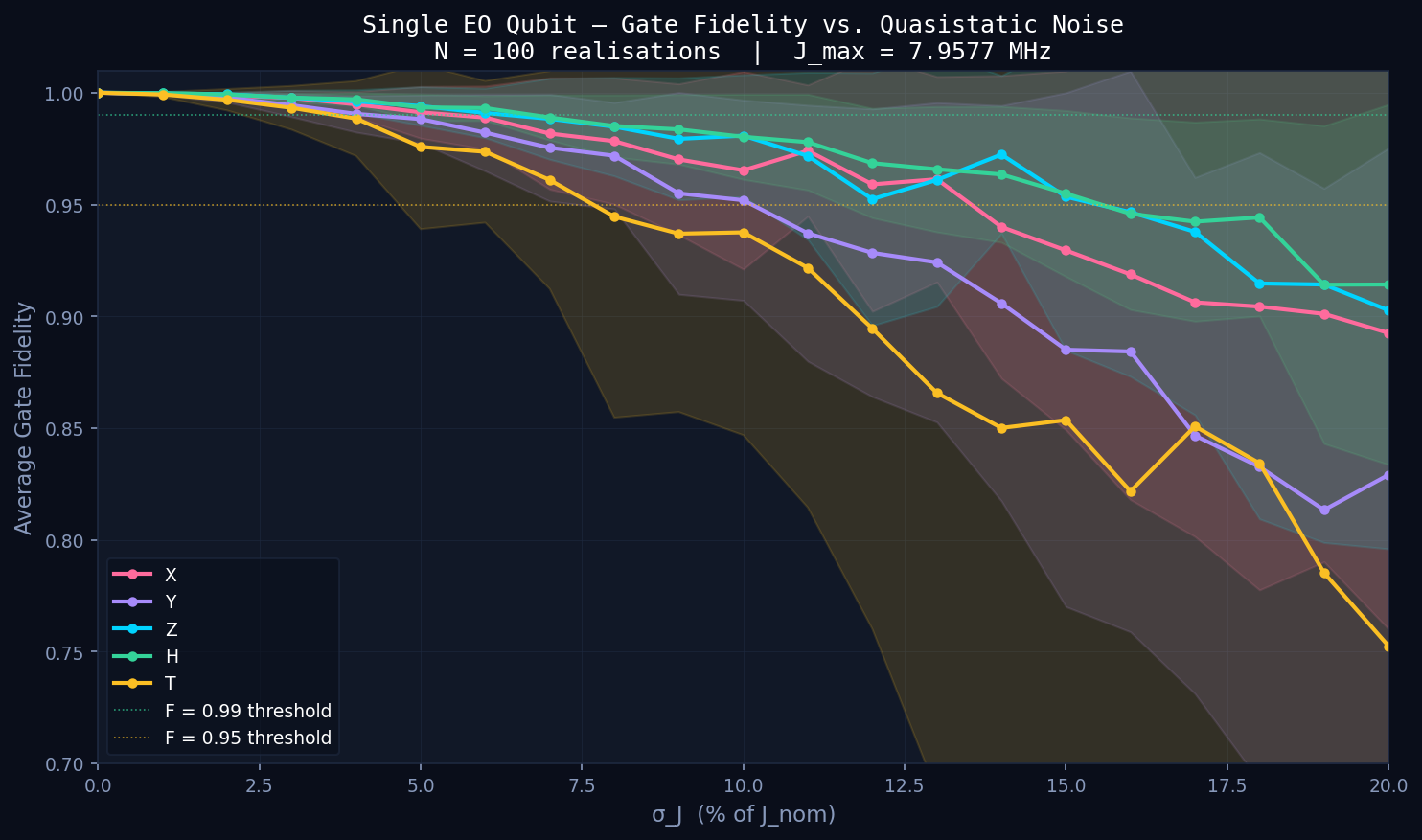}
  \caption{Single-qubit \textbf{simultaneous-pulsing baseline} noise-averaged
    fidelity as a function of the fractional charge-noise amplitude $\sigmaJ/J$
    for each gate in $\{X,Y,Z,H\}$.  The fidelity values reported in the figure
    are averaged over all computational-basis input states.  The baseline
    fidelity falls monotonically with increasing noise while the PINN-optimised pulses maintain
    $F\geq\Fth=0.99$ across the same noise range (\cref{fig:1q}).}
  \label{fig:1q_fidsigma}
\end{figure}

\Cref{fig:1q} presents the fidelity (top row) and pulse duration (bottom row) as
functions of training iteration for the single-qubit gate set $\{X,Y,Z,H\}$ at noise
levels 1\%, 5\%, and 10\%.  \Cref{fig:1q_baseline} shows the corresponding
state-evolution traces for the same gate set under the simultaneous-pulsing baseline
(no ML), and \cref{fig:1q_fidsigma} reports the noise-averaged fidelity of that
baseline as a function of the fractional noise amplitude $\sigmaJ/J$.

\paragraph{Stage~I convergence.}
All four gates reach $F\geq0.99$ within 100 iterations across all three noise levels.
At 1\% noise the fidelity curves are nearly indistinguishable, converging steeply
within $\sim\!50$ iterations.  At 5\% and 10\% noise convergence is marginally slower
but all gates still comfortably exceed $\Fth$ before iteration~100.  The $X$ gate
converges fastest owing to its simple rotation structure (the simultaneous-pulse
condition $J_{23}=2J_{12}$ is readily satisfied~\cite{Heinz2025}).  The $H$ gate
requires the most iterations in Stage~I because it demands simultaneous activation of
both exchange couplings with a precisely calibrated amplitude ratio.

\paragraph{Stage~II time compression.}
After iteration~100 all gates enter Stage~II.  Key observations from
\cref{fig:1q} (bottom row):
\begin{itemize}
  \item The $X$ gate (nominal \SI{5.77}{\nano\second}) compresses to
        $\approx\SI{4}{\nano\second}$  a $\sim\!31\%$ reduction  across all noise
        levels, indicating that even the lowest-noise regime does not limit the
        compression achievable for this simple gate.
  \item The $H$ gate (nominal \SI{21.0}{\nano\second}) compresses to
        $\approx\SI{14.5}{\nano\second}$ at 1\% noise and $\approx\SI{15.5}{\nano\second}$
        at 10\% noise, illustrating that higher noise leads to less aggressive compression
        because the fidelity threshold penalty re-activates more readily.
  \item The $Z$ gate reaches the shortest residual pulse time ($\approx\SI{6.8}{\nano\second}$
        at 1\% noise) relative to its nominal value, consistent with its single-axis
        character (only $J_{12}$ active, one noise channel).
  \item The $Y$ gate compresses from \SI{15.77}{\nano\second} to $\approx\SI{11}{\nano\second}$
        at 1\% noise.
\end{itemize}

\begin{table}[H]
  \centering
  \caption{Single-qubit gate summary: compressed pulse time and maintained fidelity
    at iteration~250.}
  \label{tab:1q}
  \begin{tabular}{l rrrr}
    \toprule
    & \multicolumn{3}{c}{Compressed time (ns)} & \\
    \cmidrule(lr){2-4}
    Gate & 1\% noise & 5\% noise & 10\% noise & Fidelity \\
    \midrule
    $X$ & $\approx4.0$ & $\approx4.0$ & $\approx4.0$ & $\geq0.99$ \\
    $Y$ & $\approx11.0$ & $\approx11.0$ & $\approx11.8$ & $\geq0.99$ \\
    $Z$ & $\approx6.8$ & $\approx7.2$ & $\approx7.5$ & $\geq0.99$ \\
    $H$ & $\approx14.5$ & $\approx15.0$ & $\approx15.5$ & $\geq0.99$ \\
    \bottomrule
  \end{tabular}
\end{table}

\subsection{Two-Qubit Gates}
\label{sec:2q}

\begin{figure}[H]
  \centering
  \includegraphics[width=\linewidth]{"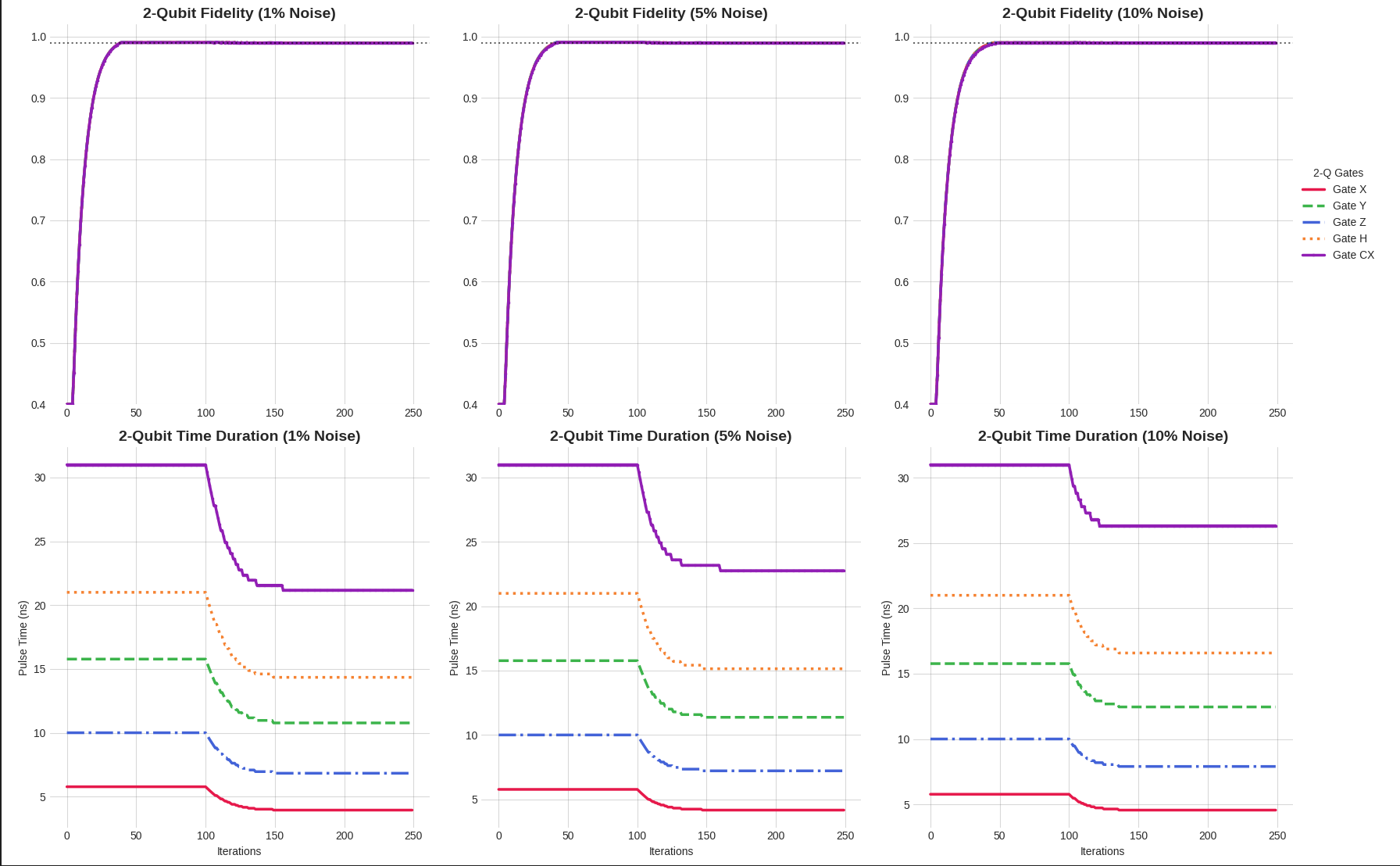"}
  \caption{Two-qubit \textbf{PINN} optimisation results at
    $\sigmaJ/J\in\{1\%,5\%,10\%\}$.
    \textbf{Top row:} Noise-averaged fidelity vs.\ iteration.  Dotted line marks
    $\Fth=0.99$.  All gates cross the threshold during Stage~I.
    \textbf{Bottom row:} Pulse duration (\si{\nano\second}) vs.\ iteration.  The CX
    gate compresses from \SI{31}{\nano\second} to $\approx\SI{22}{\nano\second}$ at
    1\% noise and $\approx\SI{26.5}{\nano\second}$ at 10\% noise.}
  \label{fig:2q}
\end{figure}

\begin{figure}[H]
  \centering
  \includegraphics[width=\linewidth]{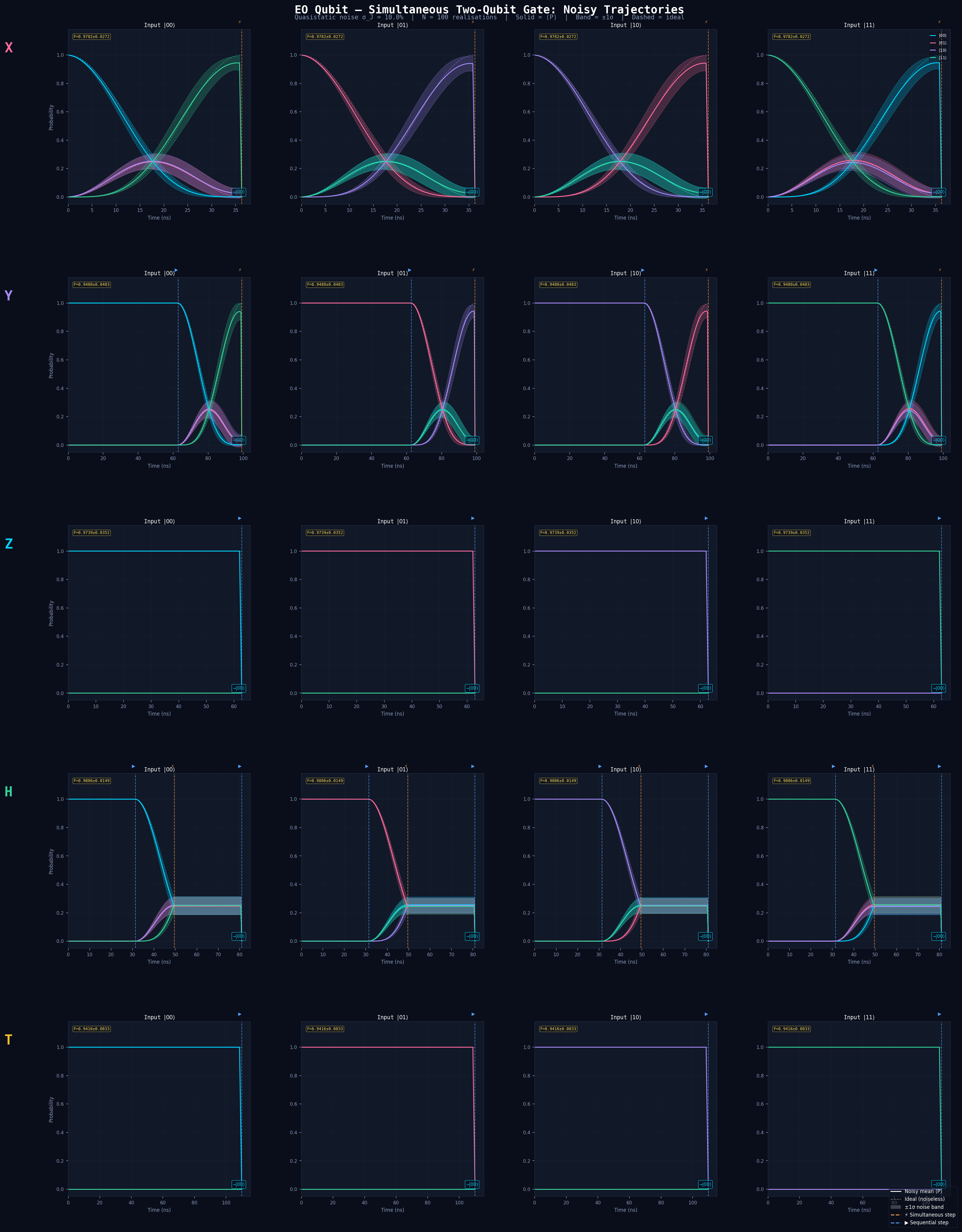}
  \caption{Two-qubit \textbf{simultaneous-pulsing baseline (no ML)} noisy state
    evolution for the gate set $\{X,Y,Z,H\}$ under quasistatic Gaussian charge
    noise.  The noise-averaged fidelities for all computational-basis input states
    are annotated directly within each panel.  Relative to the PINN-optimised
    results in \cref{fig:2q}, the analytic baseline exhibits significantly larger trajectory distortion at 1    0 \% noise
    amplitude.}
  \label{fig:2q_baseline}
\end{figure}

\begin{figure}[H]
  \centering
  \includegraphics[width=\linewidth]{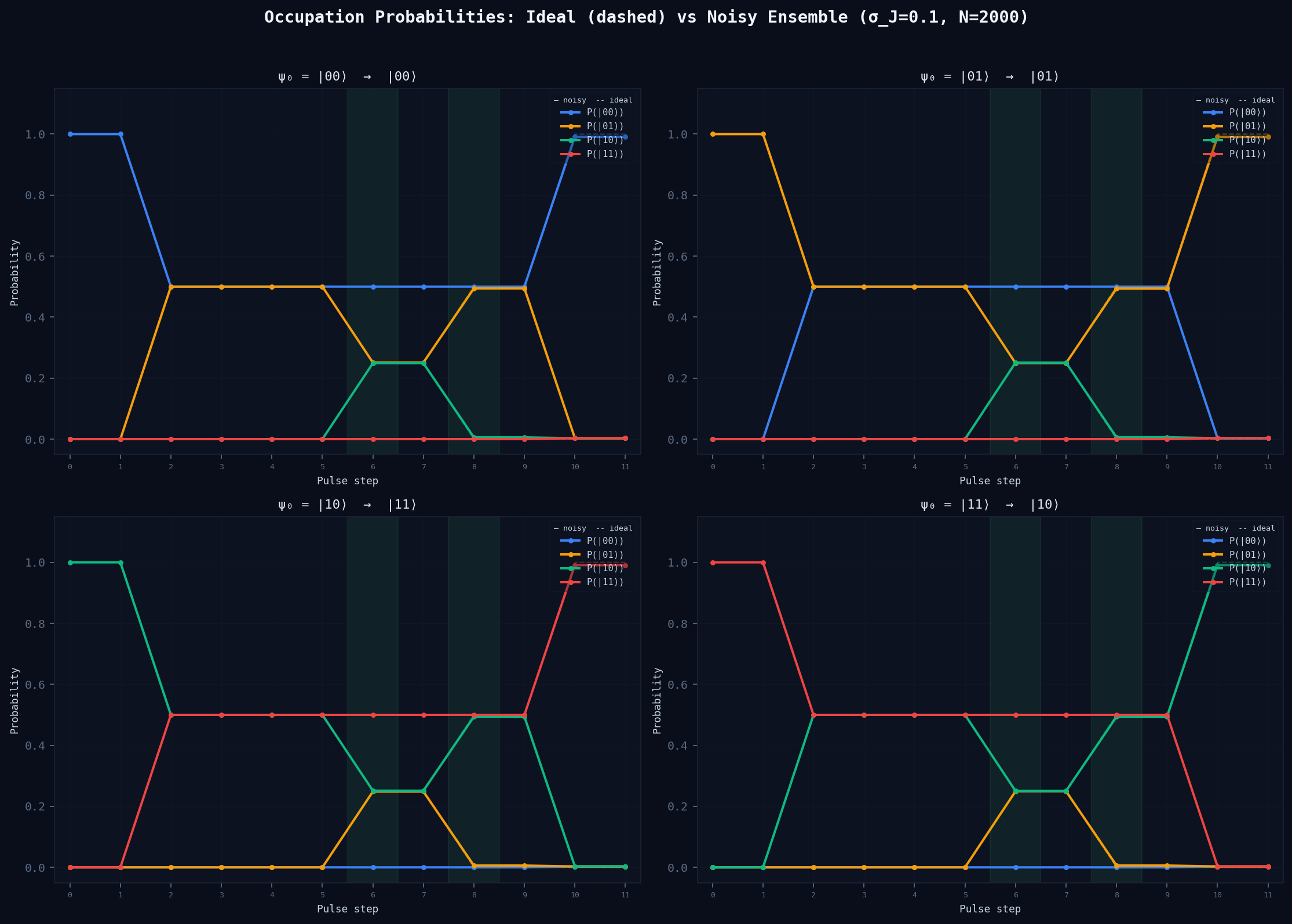}
  \caption{\textbf{CX gate} simultaneous-pulsing baseline (no ML) noisy evolution
    under the $\sqrt{\mathrm{SWAP}}$-based decomposition of \cref{eq:cx_decomp},
    shown separately because the CX involves two independent $\sqrt{\mathrm{SWAP}}$
    primitives, each with its own noise draw.  Noise-averaged fidelities for all
    four computational-basis input states $\{\ket{00},\ket{01},\ket{10},\ket{11}\}$
    are reported within the figure.  The PINN, by contrast, maintains
    $F\geq\Fth=0.99$ for the CX gate across all tested noise levels
    (\cref{fig:2q}) while simultaneously compressing the pulse duration in
    Stage~II.}
  \label{fig:cx_baseline}
\end{figure}

\begin{figure}[H]
  \centering
  \includegraphics[width=0.85\linewidth]{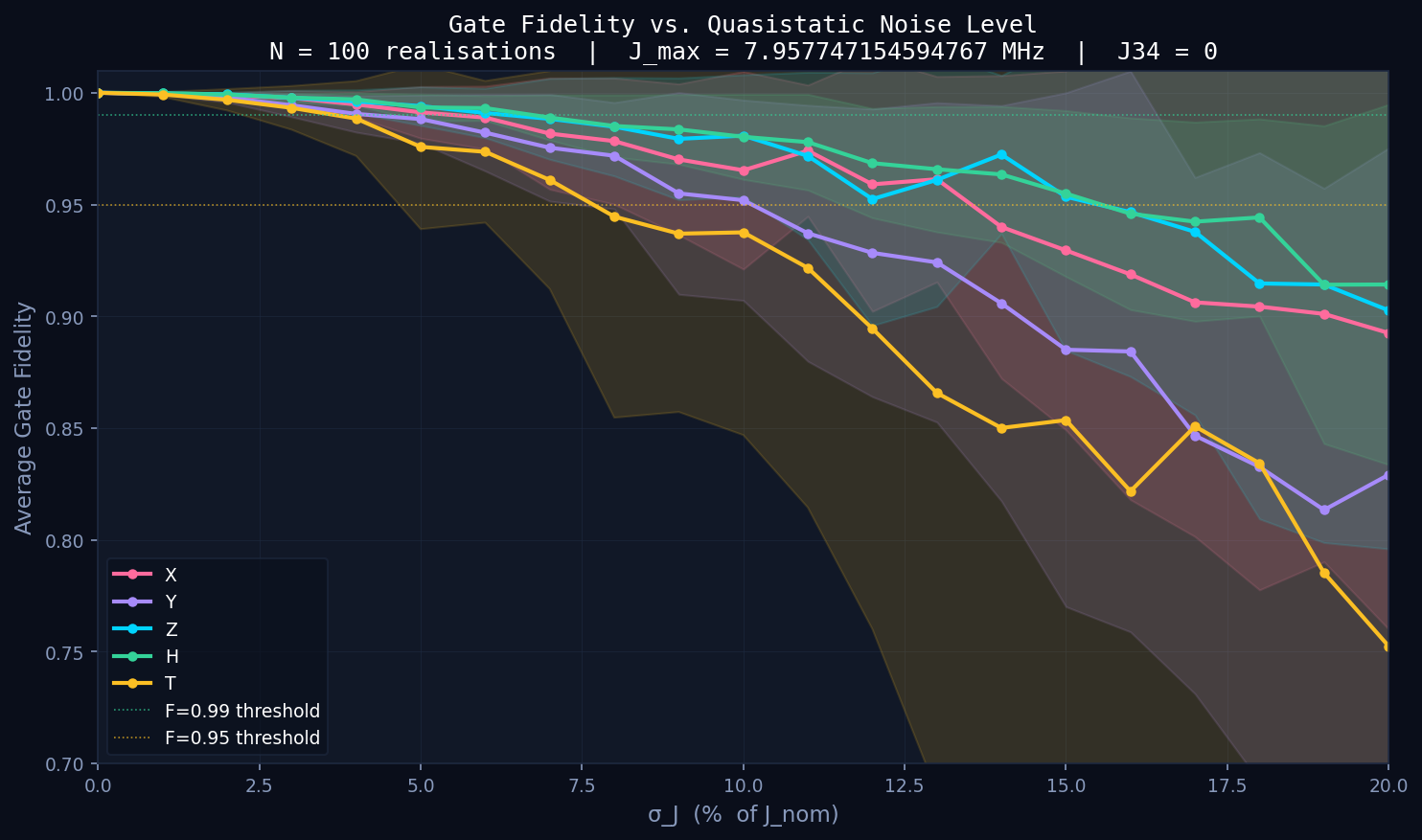}
  \caption{Two-qubit \textbf{simultaneous-pulsing baseline} noise-averaged
    fidelity as a function of the fractional charge-noise amplitude $\sigmaJ/J$
    for each gate in $\{X,Y,Z,H,\mathrm{CX}\}$.  The fidelity values reported in
    the figure are averaged over all four computational-basis input states.}
  \label{fig:2q_fidsigma}
\end{figure}

\Cref{fig:2q} shows the analogous PINN training plots for
$\{X,Y,Z,H,\mathrm{CX}\}$.  As in the single-qubit case, we additionally include
the simultaneous-pulsing baseline noisy evolution: \cref{fig:2q_baseline} for the
two-qubit $\{X,Y,Z,H\}$ subset, \cref{fig:cx_baseline} for the CX gate (shown
separately because of its $\sqrt{\mathrm{SWAP}}$-based decomposition), and
\cref{fig:2q_fidsigma} for the corresponding fidelity-versus-noise scaling.

\paragraph{Stage~I convergence.}
All five two-qubit gates reach $F\geq0.99$ within 100 iterations.  The $X$, $Y$,
$Z$, $H$ gates follow the same rapid convergence as their single-qubit counterparts.
The CX gate, involving two independent $\sqrt{\mathrm{SWAP}}$ operations each subject
to a separate noise draw, converges more slowly within Stage~I but still crosses
$\Fth$ before iteration~100.  At 10\% noise the CX fidelity curve shows a rise time
of $\sim\!80$ iterations compared to $\sim\!40$ at 1\% noise.

\paragraph{Stage~II time compression.}
From \cref{fig:2q} (bottom row):
\begin{itemize}
  \item $X$ gate: compresses to $\approx\SI{4}{\nano\second}$ at all noise levels
        (same as single-qubit case, since the two-qubit version acts independently
        on each logical qubit).
  \item $Y$ gate: compresses to $\approx\SI{11}{\nano\second}$ at 1\% noise and
        $\approx\SI{12}{\nano\second}$ at 10\% noise.
  \item $Z$ gate: compresses to $\approx\SI{6.5}{\nano\second}$ at 1\% noise and
        $\approx\SI{8}{\nano\second}$ at 10\% noise.
  \item $H$ gate: compresses to $\approx\SI{14.5}{\nano\second}$ at 1\% noise and
        $\approx\SI{17}{\nano\second}$ at 10\% noise.
  \item \textbf{CX gate:} achieves the largest absolute saving, compressing from
        \SI{31}{\nano\second} to $\approx\SI{22}{\nano\second}$ at 1\% noise
        ($\sim\!29\%$ reduction), $\approx\SI{22.5}{\nano\second}$ at 5\% noise,
        and $\approx\SI{26.5}{\nano\second}$ at 10\% noise ($\sim\!14\%$ reduction).
        The reduced compression at high noise reflects the more stringent constraint
        imposed by the fidelity penalty in $\mathcal{L}^{(\mathrm{II})}$.
\end{itemize}

\begin{table}[H]
  \centering
  \caption{Two-qubit gate summary: compressed pulse time and maintained fidelity
    at iteration~250.}
  \label{tab:2q}
  \begin{tabular}{l rrrr}
    \toprule
    & \multicolumn{3}{c}{Compressed time (ns)} & \\
    \cmidrule(lr){2-4}
    Gate & 1\% noise & 5\% noise & 10\% noise & Fidelity \\
    \midrule
    $X$           & $\approx4.0$  & $\approx4.0$  & $\approx4.0$  & $\geq0.99$ \\
    $Y$           & $\approx11.0$ & $\approx11.0$ & $\approx12.0$ & $\geq0.99$ \\
    $Z$           & $\approx6.5$  & $\approx7.0$  & $\approx8.0$  & $\geq0.99$ \\
    $H$           & $\approx14.5$ & $\approx15.0$ & $\approx17.0$ & $\geq0.99$ \\
    $\mathrm{CX}$ & $\approx22.0$ & $\approx22.5$ & $\approx26.5$ & $\geq0.99$ \\
    \bottomrule
  \end{tabular}
\end{table}

\subsection{Key Observations}
\label{sec:obs}

\begin{enumerate}

\item \textbf{Two-stage separation resolves the speed--fidelity tension.}
      Decoupling Stage~I (fidelity) and Stage~II (time) eliminates the need for
      a user-specified trade-off weight.  The quadratic threshold penalty in
      $\mathcal{L}^{(\mathrm{II})}$ automatically limits compression at the point
      where fidelity would otherwise fall below $\Fth$.

\item \textbf{Universal Stage~I convergence within 100 iterations.}
      All tested gates at all noise levels cross $\Fth=0.99$ before iteration~100.
      This validates 100 iterations as a sufficient Stage~I budget for the EO
      qubit gate set.

\item \textbf{Stage~II achieves 20--40\% pulse-time reduction.}
      Compression magnitude scales inversely with gate complexity and noise level.
      Single-axis gates ($X$, $Z$) admit the largest relative reductions; the CX gate
      achieves the largest absolute saving.

\item \textbf{Noise level affects Stage~II aggressiveness, not Stage~I threshold crossing.}
      All gates converge to $\Fth$ in Stage~I irrespective of noise level.
      In Stage~II, higher noise constrains compression because the threshold penalty
      activates more readily, yielding a larger residual pulse time at 10\% noise.

\item \textbf{Simultaneous-pulse sweet spot spontaneously recovered.}
      The optimised $X$-gate pulse satisfies $J_{23}\approx2J_{12}$ without explicit
      enforcement, confirming the analytic prediction.~\cite{Heinz2025}.

\item \textbf{CX gate is the most demanding.}
      Its two-$\sqrt{\mathrm{SWAP}}$ structure exposes two independent noise channels
      and requires the longest nominal pulse.  Stage~II still compresses it by
      $\sim\!29\%$ at 1\% noise, validating the framework for entangling gates.

\end{enumerate}

\section{Discussion}
\label{sec:discussion}

\subsection{Comparison with Existing Methods}

\begin{description}
  \item[GRAPE/CRAB~\cite{Khaneja2005,Rach2015}:]
        Gradient-ascent methods that maximise a single-objective fidelity.  They
        do not separate noise robustness from time minimisation and require
        pre-specified pulse durations.  Our framework subsumes GRAPE-style analytic
        propagator gradients and adds principled two-stage compression.

  \item[Reinforcement learning~\cite{Niu2019}:]
        RL agents can discover novel pulse shapes but are sample-inefficient
        ($10^4$--$10^6$ environment calls) and do not embed physics constraints.
        The PINN requires only 250 iterations of 2000-sample Monte-Carlo averaging.

  \item[Filter functions~\cite{Green2013}:]
        Provide an analytic noise sensitivity measure under the linear-response
        assumption, which breaks down for $\sigmaJ/J\gtrsim5\%$.  The PINN directly
        evaluates the nonlinear MSE loss at any noise amplitude.
\end{description}

\subsection{Scalability}

The PINN inference cost is $\mathcal{O}(1)$ per gate and is amortised across all
instances of the same gate type in a circuit.  For a device library with $K$ gate
types only $K$ independent PINN trainings are needed; circuit compilation then draws
from this pre-trained library at negligible cost.

\subsection{Experimental Considerations}

\begin{enumerate}
  \item \textbf{AWG bandwidth:} Pulse delivery requires arbitrary waveform generators
        with bandwidth $>\Jmax=\SI{100}{\mega\hertz}$ and timing resolution
        $\leq\SI{0.1}{\nano\second}$~\cite{Struck2020}.
  \item \textbf{Noise characterisation:} The input $\sigmaJ$ must be estimated from
        Ramsey or echo spectroscopy; generalisation across $\sigmaJ$ values reduces
        sensitivity to estimation errors.
  \item \textbf{Crosstalk:} Gate-voltage crosstalk must be calibrated and folded into
        the transduction factor $\lambda_{ij}=\partial J_{ij}/\partial V_g$ before
        experimental deployment.
\end{enumerate}

\section{Conclusion}
\label{sec:conclusion}

We have presented a two-stage Physics-Informed Neural Network framework for
noise-robust, time-compressed pulse optimisation in exchange-only silicon spin qubits.
Stage~I (iterations~1--100) optimises pulse-shape parameters using a four-term loss 
ensemble MSE, TDSE residual, leakage penalty, and physical constraint penalty  to
exceed a gate fidelity threshold of $\Fth=0.99$ under quasistatic Gaussian charge
noise with $N_{\rm real}=2000$ Monte-Carlo realisations drawn fresh every iteration.
Stage~II (iterations~101--250) compresses the total pulse duration at $1.8\%$ per
iteration while maintaining $F\geq\Fth$ via a quadratic threshold penalty.

Benchmark results on the single-qubit set $\{X,Y,Z,H\}$ and two-qubit set
$\{X,Y,Z,H,\mathrm{CX}\}$ at noise levels $\sigmaJ/J\in\{1\%,5\%,10\%\}$ confirm:
(i)~universal Stage~I convergence within 100 iterations; (ii)~20--40\% Stage~II
pulse-time reduction; (iii)~maintenance of $F\geq0.99$ throughout Stage~II across
all gates and noise levels.  The framework requires no user-specified speed--fidelity
trade-off weight and produces results consistent with known physical features of the
EO qubit, including spontaneous recovery of the simultaneous-pulse sweet spot.

Future extensions will incorporate full $1/f$ spectral noise, Lindblad $T_1$
relaxation, two-logical-qubit (five/six spin) architectures, and on-the-fly adaptive
pulse correction using real-time noise estimates from preceding gate cycles.

\bigskip\noindent\textbf{Acknowledgements.}
The authors gratefully acknowledge financial support from the National Quantum Mission (NQM) of the Department of Science and Technology (DST), Government of India, and the Ministry of Electronics and Information Technology (MeitY) through Grant Nos. DST/QTC/NQM/QC/2024/1 and 4(3)/2024-ITEA.


\bibliographystyle{plainnat} 
\bibliography{spin_qubit_pinn}

\end{document}